\documentclass{aac}

\usepackage[numbers]{natbib}
\usepackage{rotating}

\usepackage{xcolor}
\usepackage{graphicx}
\begin{document}

\title{Analytical Solution to Electron Beam Envelope Evolution in a Plasma Wakefield}

\author[aff1]{Jiayang Yan\corref{cor1}}
\author[aff2]{Audrey Farrell}
\author[aff3]{L\'{\i}gia Diana Amorim}
\author[aff1]{Navid Vafaei-Najafabadi }

\affil[aff1]{Stony Brook University}
\affil[aff2]{University of California, Los Angeles
}
\affil[aff3]{ Lawrence Berkeley National Laboratory
}
\corresp[cor1]{jiayang.yan@stonybrook.edu}

\maketitle

\begin{abstract}
 Beam-induced ionization injection (B-III) is currently being explored as a method for injecting an electron beam with a controlled density profile into a plasma wakefield accelerator (PWFA).  This process is initiated by the fields of an unmatched drive beam where the slice envelope reaches its minimum value, the ``pinch”.
To control the injected beam's qualities, it is crucial to study the beam-slice envelope oscillations, especially size and the location of the pinch. 
In this proceeding, an ansatz based on the harmonic motion is proposed to find the analytical solution to beam-slice envelope evolution in the nonlinear regime. The size of the pinch is then found through the application of energy conservation in the transverse direction. 
The resulting analytical expressions are shown to be in good agreement with numerical solutions.

\end{abstract}

%
\section{INTRODUCTION}

Beam-driven plasma wakefield accelerators (PWFA) use electron drivers to create plasma cavities where the acceleration gradient can reach tens of GeV/m \cite{bib1}. In the so-called nonlinear regime \cite{bib2}, the plasma electrons are repelled from the axis by the drive electron beam due to the Coulomb force, while heavier plasma ions are left behind the driver. The stationary ions provide focusing forces that attract plasma electrons back to the axis, forming a bubble-like structure. A trailing electron beam located at the back of the bubble gains energy from the generated electromagnetic field, the wakefield. The wakefield provides a transverse focusing force in which the electrons execute an oscillatory motion, called the betatron oscillation, around the axis.

One method that can be used to form and inject the trailing beam inside the ion cavity is called Beam-Induced Ionization Injection (B-III) \cite{bib4}, where the plasma is comprised of a low-ionization-threshold (LIT) gas (high concentration) plus a high-ionization-threshold (HIT) impurity component (low concentration) and the driver is unmatched with the plasma. During the wake excitation, as the initial drive beam space charge field does not go above the high ionization threshold, only the LIT gas is ionized to form the wake structure. Then because of the betatron oscillation, the envelope of each beam slice oscillates in time, and so does the beam field. As the transverse size of the beam slice reduces to its minimum (pinch), the beam field increases and exceeds the high ionization threshold. As a result, the HIT impurity can be ionized and becomes the source for the injected trailing beam. 

To control the ionization and injection process, which is initiated around the pinch, it is required to know the evolution function of beam slice envelope along the propagation direction as well as the size of the pinch. In this proceeding, an ansatz is proposed to find the analytical solution to beam-slice envelope oscillation in the nonlinear regime. 
This ansatz is based on the application of energy conservation in the transverse direction. The solution is first developed for the case of an electron beam with constant energy and is then generalized to the case of an accelerating/decelerating beam. 
The analytical results are shown to be in good agreement with numerical solutions.

It is important to note that while this envelope oscillation calculation is intended here for applications in the case of a PWFA, the results are general and can apply to the case of a trailing electron beam in a laser driven wakefield accelerator (LWFA) as well. 

\section{EVOLUTION OF BEAM-SLICE ENVELOPE}

Evolution of a beam-slice envelope $ \sigma_{r} (z)$ undergoing betatron oscillation is expressed as a function of propagation distance $z$ by \cite{bib3}
\begin{equation}
\frac{d^{2} \sigma_{r} (z)}{d z^{2}} +k_{\beta}^{2} \sigma_{r}(z)-\frac{\epsilon_{N}^{2}}{\gamma^{2} \sigma_{r}^{3}(z)} = 0, \label{eqn:envelope}
\end{equation}
where $k_\beta =\frac{\omega_p}{c} \sqrt{\frac{1}{2\gamma(z)}}$ is the betatron frequency, $\omega_p = \sqrt{\frac{e^2n_0}{m_e\epsilon_0}}$ is plasma frequency, $\gamma $ is the Lorentz gamma factor, $n_0$ is plasma density, $e$, $m_e$ are electron charge and mass respectively, and $\epsilon_{N}$ is the normalized beam emittance. 
In the equation, $F_+ = k_{\beta}^{2} \sigma_{r}(z)$ is the focusing force of the wakefield, and $F_- =- \frac{\epsilon_{N}^{2}}{\gamma^{2} \sigma_{r}^{3}(z)}$ works as an effective defocusing force. It is well known that the solution to Eq.~(\ref{eqn:envelope}) is a betatron oscillation with the wavenumber $k_\beta$ and a minimum size that is called the ``pinch". As the exact solution of Eq.~(\ref{eqn:envelope}) is too complicated to be solved, an ansatz is proposed to describe the betatron motion as a harmonic oscillation with a minimum size (pinch) of $ \Delta\sigma_r$. 

\begin{equation}
    \sigma_r (z) = |(\sigma_{r_0} - \Delta\sigma_r)\cos(k_\beta z)|+\Delta\sigma_r. \label{eqn:ansatz}
\end{equation}
Here, $\sigma_{r_0} $ is the initial transverse beam size, which is known. 
The consequence of this ansatz is that $F_-$ can be written in terms of $ \Delta\sigma_r$ in the same frequency as $F_+$ is written in terms of $\sigma_r$. 

The next step is to find the value of the pinch size, $\Delta\sigma_r$, which is determined by balancing the energy that would be associated with each force term in an oscillation period.
\\Consider a half cycle, 
\begin{equation}
   W_+ = \int^{\sigma_{r_{0}}} _{\Delta\sigma_r} F_+ (\sigma_r) d \sigma_r = {\omega_p^2\over 4{ c^2\gamma}} \left(\sigma_{r_0}^2  - \Delta \sigma_r^2 \right),\label{enq:w+}
\end{equation}

\begin{equation}
    W_- =  \int^{\sigma_{r_{0}}} _{\Delta\sigma_r} F_- (\sigma_r) d \sigma_r  = \frac{\epsilon_N^2}{2\gamma^2} \left({1\over \sigma_{r_0}^2 }- {1\over \Delta\sigma_r^2 }\right). \label{enq:w-}
\end{equation}
The energy is conserved in the half cycle, so $ \Delta\sigma_r$ can be found from $  W_+ +W_- = 0 $:
\begin{equation}
    \Delta\sigma _r^2 = {1\over 2}\left[\sigma_{r_{0}}^2 + \frac{2\epsilon_N^2 c^2}{\omega_p^2 \gamma\sigma_{r_{0}} }\pm \sqrt{\left(\sigma_{r_{0}}^2 + \frac{2\epsilon_N^2 c^2}{\omega_p^2 \gamma\sigma_{r_{0}} }\right)^2 - 4 \frac{2\epsilon_N^2 c^2}{\omega_p^2 \gamma }}\right] = \frac{2 \epsilon_N^2 c^2}{\omega_p^2 \gamma \sigma_{r_{0}}^2},
\end{equation}
\begin{equation}
    \Delta\sigma_r=\sqrt{{2\over \gamma }}\frac{ \epsilon_N c}{\omega_p \sigma_{r_0}}\label{enq:sol}.
\end{equation}
\begin{figure}
\centerline{\includegraphics[width=4.5 in]{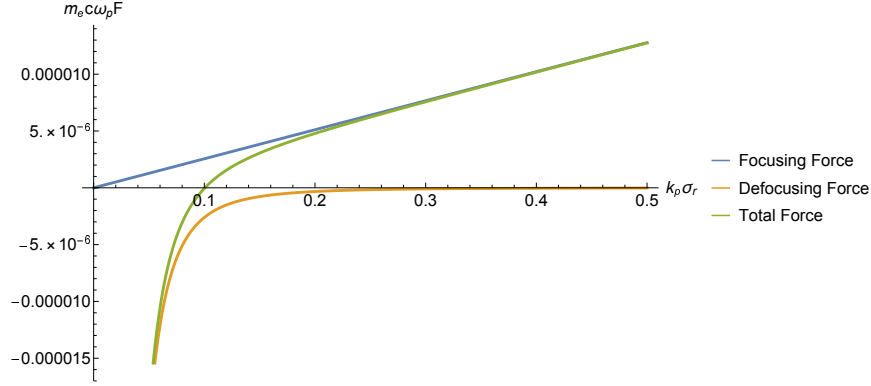}}
\caption{Forces in Eq.~(\ref{eqn:envelope}), where $\gamma = 19570.5$ (10 GeV), $k_p\sigma_{r_0} =0.5$, and
$k_p \epsilon_N = 1$.}
\label{fig:force}
\end{figure}
A conceptually simple understanding of $\Delta\sigma_r$ can emerge by studying the behavior of the total force $F_++F_-$ in the limiting cases of $\sigma_r$.
From their mathematical expressions, we expect the total force $F_+ + F_-$ to be dominated by $F_+$ at a large $\sigma_r $, where $-\frac{1}{\gamma^{2} \sigma_{r}^{3}(z)} \rightarrow 0$. On the other hand, $F_-$ is expected to dominate when $\sigma_r $ is small and $\sigma_r  \ll \left(\frac{1}{\gamma}\right)^{1/4}$. 
The plot of the total force as well as $F_+$ and $F_-$ as functions of transverse position are shown in Fig. \ref{fig:force} using normalized units. It is clearly seen here that the total force (in green) follows $F_+$ (in blue) when $\sigma_{r}$ is large (i.e., $\sigma_{r}\ge 0.15 $ in the figure); on the contrary, it follows $F_-$ (in orange) when $\sigma_{r}$ is small (i.e., $\sigma_{r}\le 0.05 $). The equilibrium position $\sigma_{r_{eq}} $, defined as the position where the forces balance, $F_+ = F_-$, is given by the plasma matched spot size
\begin{equation}
     \sigma_{r_{eq}} = \left({\epsilon_N \over \gamma k_\beta}\right)^{1\over 2}. \label{eqn:req}
\end{equation}
Because different forces dominate the interaction on either sides of $\sigma_{r_{eq}}$, a simplified physical model for the betatron oscillation may be prescribed as follows: only $F_+$ acts on the electrons when $\sigma_{r_0}> \sigma_r > \sigma_{r_{eq}} $ and only $F_-$ acts on the electrons for $\Delta \sigma_r <\sigma_r < \sigma_{r_{eq}}$, with $\sigma_{r_{eq}}$ considered as the transition point. 
This model allows for the work done by $F_+$ to be found independently of the size of the pinch, $\Delta\sigma_r$, 
\begin{equation}
   W_+^\prime = \int_{\sigma_{r_{eq}}} ^{\sigma_{r_0}} F_+ (\sigma_r) d \sigma_r = {\omega_p^2\over { c^2\gamma}} \left({\sigma_{r_0}^2 \over2} - \sqrt{\epsilon_N\over{2 \gamma}}\right). \label{enq:w+}
\end{equation}
$\Delta\sigma_r$ then can be simply interpreted as the distance through which $F_-$ needs to be exerted to cancel the energy gained by $F_+$. 
Applying the same conservation of energy criteria as discussed above, $W_+^\prime+W_-^\prime=0$, with $W_+^\prime$ defined in Eq.~(\ref{enq:w+}) and $W_-^\prime$ given by
\begin{equation}
    W_-^\prime =  \int^{\sigma_{r_{eq}}} _{\Delta\sigma_r} F_- (\sigma_r) d \sigma_r  = -\frac{\epsilon_N^2}{4 \gamma^2} \left({\sqrt{2 \gamma \over \epsilon_N}}- {2\over \Delta\sigma_r^2 }\right), \label{enq:w-}
\end{equation}
leads to the same expression for $\Delta\sigma_r$ as Eq.~(\ref{enq:sol}), which was obtained using the exact expressions.  
\begin{figure}
\centering 
\includegraphics[width=0.7\textwidth]{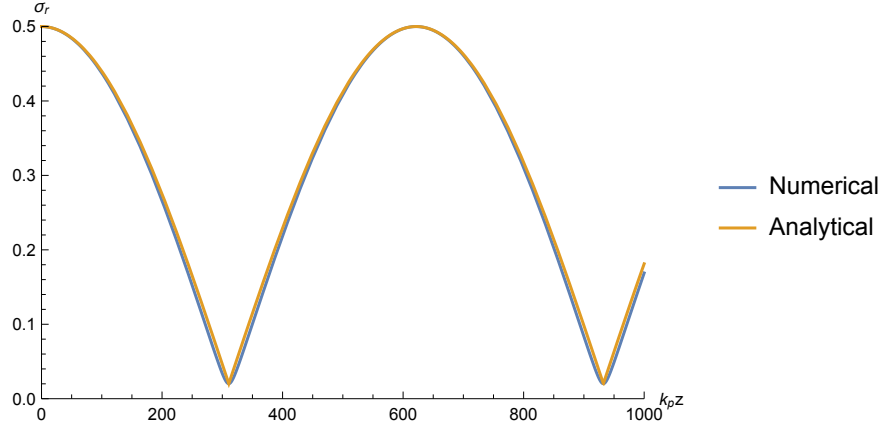}
\caption{Comparison between numerical solution and analytical ansatz, where $k_p\sigma_{r_0} =0.5$, $k_p \epsilon_N = 1$, $\gamma = 19570.5$ (10 GeV), $\Delta \sigma_r = 0.0202183$  from Eq.~(\ref{enq:sol}), numerical pinch size is 0.0202185. The relative difference is 0.01\%. The absolute difference between both solutions is in the order of $10^{-7}$. }
\label{fig:compare}
\end{figure}
Mathematically, $ \Delta\sigma_r$ represents the transverse offset between the minimum value of $\sigma_r(z)$ and z axis, which is the size of the pinch.

The analytical ansatz in Eq.~(\ref{eqn:ansatz})
 is compared with the numerical solution of Eq.~(\ref{eqn:envelope}) shown in Fig. \ref{fig:compare} for a sample set of parameters described in the figure caption. 
Based on Eq.~(\ref{enq:sol}), $\Delta\sigma_r = 0.0202183$, numerical pinch size is 0.0202185. The relative difference is $0.01\%\ll 1\%$. The absolute difference between the numerical and analytical solutions was checked for multiple cases where $10 < \gamma < 10000$, and the average difference was found to be on the order of $10^{-7}$. This indicates that the expression in Eq.~(\ref{enq:sol}) is also valid for a low energy beam.
\section{THE IMPACT OF ENERGY VARIATION}

\begin{figure}
\centerline{\includegraphics[width=5 in]{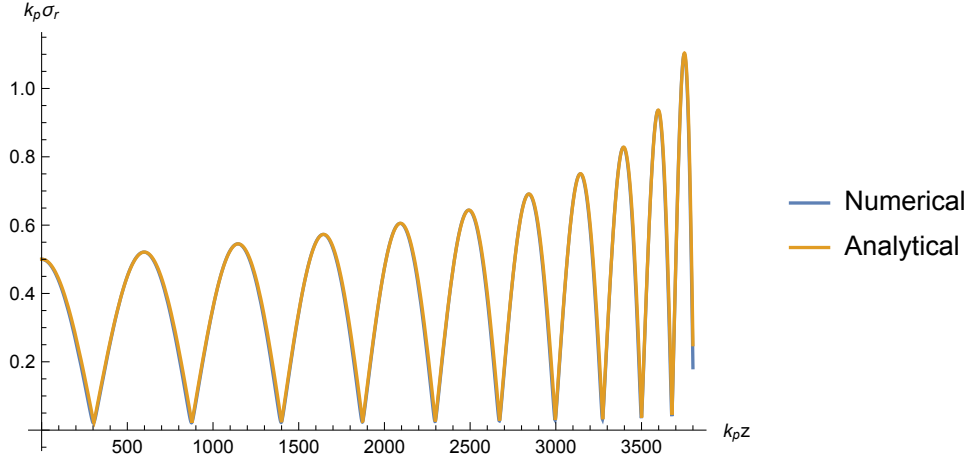}}
\caption{Comparison between numerical solution for Eq.~(\ref{eqn:ansatz2}) (blue) and analytical ansatz (orange) with acceleration, where $\gamma_0 = 19570.5 $ (10 GeV), $\gamma(z) = \gamma_0 - 5 z$, $k_p\sigma_{r_0} =0.5$, and
$k_p \epsilon_N = 1$. The analytical ansatz is calculated based on Eq.~(\ref{enq:sol2}).}
\label{fig:compare2}
\end{figure}

 Taking acceleration into account, the betatron envelope equation is modified to \cite{bib5,bib6}, 
 \begin{equation}
 \frac{d^{2} \sigma_{r} (z)}{d z^{2}} + \frac{\gamma(z)^\prime}{\gamma(z)}\sigma_r^\prime(z) +k_{\beta}^{2} \sigma_{r}(z)-\frac{\epsilon_{N}^{2}}{\gamma(z)^{2} \sigma_{r}^{3}(z)} = 0. \label{eqn:envelope2}
\end{equation}
Here, typically $\gamma(z) = \gamma_0 +E_z z$ in the normalized unit, and $E_z$ is the longitudinal electric field, which is usually a constant (in $z$ for each beam slice), because the beam velocity and the phase velocity of the wake are both nearly $c$. The impact of energy gain/loss is modelled using the approach of adiabatic damping.  Based on the simplified physical model developed in the previous section, for $\sigma_r>\sigma_{r_{eq}}$, only the impact of $F_+$ can be considered. The solution to Eq.~(\ref{eqn:envelope2}) without the $F_-$ term consists of damped harmonic oscillations, where the oscillation amplitude is modified by a factor of $\left(\frac{\gamma_0}{\gamma(z)}\right)^{1/4}$, i.e. 
$\sigma_{r_{0}} \rightarrow \sigma_{r_{0}} \left(\frac{\gamma_0}{\gamma(z)}\right)^{1/4}$ \cite{bib7}.
Similar modifications for the amplitude and the phase of the ansatz [Eq.~(\ref{eqn:ansatz})] result in the following expression, where the effect of $F_-$ in Eq.~(\ref{eqn:envelope2}) is once again taken into account using the parameter $\Delta\sigma_r$:

\begin{equation}
    \sigma_r (z) = \left| \left[\sigma_{r_0} \left(\frac{\gamma
    _0}{\gamma(z)}\right)^{1/4} - \Delta\sigma_r(z)\right]  \cos\left(\int k_\beta(z) dz\right)\right|+\Delta\sigma_r(z). \label{eqn:ansatz2}
\end{equation}
Here $\Delta\sigma_r(z)$ becomes a function of $z$. The modification of $ \Delta\sigma_r$ is likewise carried out by $\sigma_{r_{0}} \rightarrow \sigma_{r_{0}} \left(\frac{\gamma_0}{\gamma(z)}\right)^{1/4}$, making $ \Delta\sigma_r(z)$ vary from one cycle to the next as $\gamma(z)$ changes,

\begin{equation}
    \Delta\sigma_r(z)=\sqrt{{2\over \gamma(z) }}\frac{ \epsilon_N c}{\omega_p \sigma_{r_0}\left(\frac{\gamma_0}{\gamma(z)}\right)^{1/4}}=\left({4  \over \gamma(z) \gamma_0 }\right)^{1/4}\frac{ \epsilon_N c}{\omega_p \sigma_{r_0}}.\label{enq:sol2}
\end{equation}
The modified analytical solution given by Eq.~(\ref{eqn:ansatz2}) is compared with the numerical solution for Eq.~(\ref{eqn:envelope2}) in Fig. \ref{fig:compare2}. The initial energy of the drive beam is 10 GeV and the deceleration field is -5 in the normalized units, resulting in $\gamma(z) = \gamma_0 - 5 z$. This deceleration field is chosen artificially so that the trend is more clear. The analytical solution closely tracks the numerical solution, and strong agreement is observed between the two in the description of the amplitude, frequency and the pinch size. 
\section{CONCLUSION}

In conclusion, an analytical solution for the betatron envelope oscillation was developed based on an ansatz that the solution consists of a bound harmonic oscillation. The minimum transverse size of the envelope, called the pinch, was calculated using an energy conservation argument between the two force terms in the envelope oscillation. The variation of particle energy was taken into account using the adiabatic damping of the betatron envelope model. These analytical solutions show excellent agreement with the numerical solutions to the envelope equations. The general solution $\sigma_r (z) = \left|\left[\sigma_{r_0}\left(\frac{\gamma(z)}{\gamma_0}\right)^{1/4} - \Delta\sigma_r(z)\right]  \cos\left(\int k_\beta(z) dz\right)\right|+\Delta\sigma_r(z)$, where $\Delta\sigma_r(z)=\sqrt{{2\over \gamma(z) }}\frac{ \epsilon_N c}{\omega_p \sigma_{r_0}(\frac{\gamma_0}{\gamma(z)})^{1/4}}$, gives us a valuable tool for accurately predicting the size and the location of the pinch (along $z$), which in turn allows for the calculation of the position and the magnitude of the maximum radial electric field. This information is invaluable for the B-III method as it will enable the precise positioning of the impurity elements. Alternatively, in methods that attempt to avoid additional ionization of the impurity, it will give an insight into the position where the presence of the impurity should be avoided. The size of the pinch will determine the size of the ionization region, which can be used in the future to calculate the quality of the generated electron beam in B-III.  

\section{ACKNOWLEDGMENTS}
We acknowledge the support by U.S. Department of Energy, Office of High Energy Physics (HEP) program under Award No. DE-SC-0014043 and resources of NERSC facility, operated under Contract No. DE-AC02-5CH11231.


\nocite{*}
\bibliographystyle{aac}%
\bibliography{aac2020}%



\end{document}